# INCORPORATING COMPETITION INTO DUAL ACCESSIBILITY ASSESSMENT: THE COMPETITIVE EQUILIBRIUM METHOD

**André Borgato Morelli\***
andre.morelli@usp.br
**André Luiz Cunha**
University of Sao Paulo
São Carlos School of Engineering

**ABSTRACT**
This study introduces a modified version of the Balancing Time accessibility metric, aiming to incorporate competition among users into the original process. The creation of this approach was motivated by the limitations of the previous method in accurately representing the competitive dynamics within transportation systems. The updated method, termed Competitive Equilibrium, achieves this by adjusting available opportunities continuously based on the presence of competitors, thereby effectively reflecting the impact of competition, and distinguishing between areas based on their demand and supply balance. To validate the effectiveness of Competitive Equilibrium, we tested it in three cities in São Paulo State, Brazil. Our findings show that this method is significantly more responsive to changes in employment availability and offers a more accurate depiction of competitive dynamics in urban settings. Moreover, we have made the data and computational tools used in this study publicly accessible, facilitating the reproduction of our work and its application in further research.

## 1. INTRODUCTION

In recent years, the swift urbanization of developing countries has heightened the focus on urban accessibility in the fields of urban planning and transportation (Boisjoly et al., 2020; Carneiro et al., 2019). The notion of potential accessibility, which evaluates an individual's ability to access various opportunities such as employment, education, health services, and leisure activities, has become a key metric for assessing the effectiveness and fairness of urban transport networks. This emphasis follows Hansen's pioneering exploration of gravitational accessibility metrics in 1959. Since then, scholars have introduced a range of metrics to better understand the factors influencing user accessibility (Barboza et al., 2021; Cheng and Bertolini, 2013; Higgs et al., 2019; Pereira, 2019; Shen, 1998).

Location-based accessibility measures are commonly categorized into three types: cumulative, gravitational, and utility-based (Geurs and van Wee, 2004). Cumulative measures count the opportunities reachable within a certain cost threshold, gravitational measures weigh opportunities by their distance from users (Hansen, 1959), and utility-based measures incorporate utility theory into accessibility analysis (Geurs and van Wee, 2004). Despite the dominance of these types, particularly cumulative and gravitational metrics, in research (Boisjoly et al., 2020; Hernandez et al., 2020; Levine et al., 2019; Pereira, 2019; Pritchard et al., 2019), the exploration of alternative metrics has continued, creating unconventional characterizations of urban environments. One such example is the work of Barboza et al. (2021), that proposed the Balancing Time metric to measure the time taken by a group of users to saturate their demand for opportunities. This metric differs from the other, more common, metrics due to it being a dual accessibility measure. In general, the common (primal) accessibility metrics focus on the number of opportunities reachable within a set timeframe. In contrast, dual accessibility metrics, such as Balancing Time, examine how long it takes a user to meet their needs for opportunities (Cui and Levinson, 2020). The definition of dual accessibility metrics is particularly useful to address the concept of 15-minute cities, which aim for essential services and opportunities to be within a 15-minute travel distance, promoting sustainable urban living.

In a 15-minute city, all essential urban services and amenities, such as work, shopping, education, healthcare, and leisure, are accessible to residents within a 15-minute trip from their homes in such a way that planners focus on the time required for users to access amenities (Allam et al., 2022a) and not other usual metrics such as level of service. This model promotes reduced reliance on cars, leading to lower traffic congestion, improved air quality, and enhanced community engagement (Allam et al., 2022b). Dual accessibility metrics can quantify how easily residents can access essential services within a given area by calculating the time required to reach various opportunities (employment, education, healthcare, etc.) rendering then a natural tool for providing a clear picture of the current state of urban accessibility towards a 15-minute city goal.

Nevertheless, although many metrics have been established, a critical challenge in the field of accessibility is accounting for competition among users for access to opportunities. One of the early examples of effort to address this problem is the work of Shen (1998) that proposed a two-stage approach to accessibility, adjusting the number of opportunities of a location by the competitive potential in the area:

$$A_i^{pop} = \sum_j O_j \cdot \frac{f(c_{ij})}{A_j^o} \qquad \text{Eq.1}$$

$$A_j^{emp} = \sum_k P_k f(c_{jk}) \qquad \text{Eq.2}$$

Where: $A_i^{pop}$ is the population's accessibility to opportunities; $A_j^o$ is the accessibility of the opportunities to the population (competitive potential); $f(c_{ij})$ is a distance decay ratio as a function of travel cost ($c_{ij}$) between the zones $i$ and $j$; $O_j$ is the number of opportunities in the area $j$; and $P_k$ is the population in zone k.

This method is also known as the Two-Step Floating Catchment Area or 2SFCA (Luo and Wang, 2003), and is widely known to incorporate competition into traditional accessibility measures. One drawback of the 2SFCA method, however, is the difficulty in interpreting its results since it yields abstract, dimensionless fractions instead of the number of reachable opportunities (Geurs and van Wee, 2004). Additionally, the 2SFCA method is, by nature, a primal metric. In contrast, dual accessibility metrics in general lack a robust method to address competition.

The Balancing Time method is purported to account for competition (Barboza et al., 2021); nonetheless, it if the method is closely analyzed, it becomes clear that it falls short in adequately addressing the competitive dynamics among users vying for the same opportunities, challenging this claim. To address this gap, our study introduces an enhanced version of the Balancing Time method that integrates competition, termed the Competitive Equilibrium method. This approach dynamically updates the availability of opportunities in response to the level of competition among users. We assess the performance of this new method against the original by applying both to various cities under different employment supply conditions, thereby gauging their sensitivity to competition for jobs.

This novel method is not directly comparable to 2SFCA since it lies in a different category of metrics. However, it is possible to infer that the application of this method brings several advantages about the 2SFCA, mainly regarding the ability to capture competition while still preserving an easily interpretable metric: the time to saturate demand. In a dense urban environment, where multiple people might be vying for the same opportunities (e.g., seats in a school, job openings), this method can provide a more realistic view of accessibility. Finally, the data used was made available[1] to facilitate the reproduction of this work. Furthermore, the computational tools created in the Python language to calculate the metrics described in this work can be accessed openly.

## 2. PROPOSED METHOD

In this section, we discuss our method for analyzing accessibility with a focus on competitiveness, employing a dual approach. We introduce the Competitive Equilibrium method as an alternative that is more attuned to competition dynamics than the Balancing Time proposed by Barboza et al. (2021). Additionally, we provide insights into the data sources used as a proof of concept for our method.

### 2.1. The Competitive Equilibrium method

The Balancing Time method was introduced to calculate the cost incurred by a given population to saturate its demand for opportunities within the transport network (Barboza et al., 2021). The authors applied this metric to determine the travel time needed for the entire economically active population of an area to access at least one employment position each, considering a saturation ratio of 1.0. However, this ratio can be adjusted as a calibration parameter.

The original method was presented as a low-complexity alternative to measuring competition for opportunities but had limitations. To illustrate this limitation, consider the hypothetical scenario in Figure 1(a), where residential zone A has 100 economically active individuals, and commercial zones I and II each have 100 jobs. If commercial zone I already satisfies the job demand of the population in zone A with a travel time of 10 minutes, the Balancing Time (BT) in zone A is 10 minutes.

Alternatively, consider the situation in Figure 1(b), where a new residential zone, B, with 100 population and identical costs to zone A, is added. Even though Zones A and B compete for employment in commercial zone I, for the Balancing Time method, commercial zone I still satisfies the job demand in zone B, resulting in a Balancing Time of 10 minutes for both residential zones. That happens because the original Balancing Time method does not account for the fact that when two traffic zones overlap for the same set of opportunities, they compete for these, and supply should be adjusted accordingly. In this case, while zone I has 200 people competing for 100 jobs, zone II has zero people employed for the same 100 jobs, leading to a competitive distortion.

---

[1] Available at https://github.com/labITS-stt-eesc/competitive_equilibrium

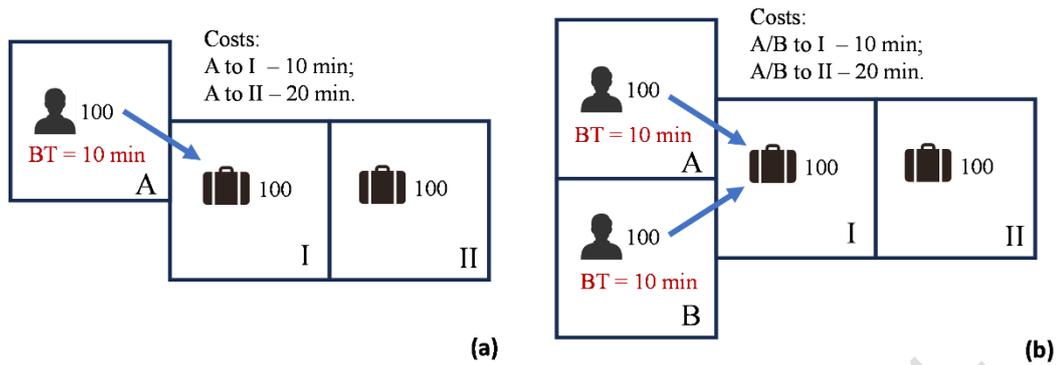

**Figure 1**: Demonstration of the calculation of the Balancing Time in a less competitive environment (a) and in an environment with additional competition. The method does not show any variation when competition increases.

To address the limitation above and capture competitive dynamics, our proposed method computes the number of potential competitors per job in each zone. We define potential competitors as the number of people competing for each job within a zone. For example, in Figure 1, zone I has 200 people competing for 100 jobs, resulting in two potential competitors per job. In contrast, zone II has zero potential competitors, as neither zone A nor B needs to seek jobs in zone II to meet their demand.

Figure 2 illustrates our proposed method on the same system as Figure 1(b). The number of potential competitors ("Compet") for each zone in the first iteration can be seen in Figure 2(a). After calculating the number of potential competitors for each zone, we determine the adjusted supply by dividing jobs in zones with multiple competitors by the number of potential competitors from the previous iteration (Figure 2(b)). If two groups compete for 100 jobs, 50 jobs will be allocated to each competing zone. As we repeatedly calculate the Balancing Time for the system, both the time and the number of competitors per zone change, leading to an iterative process until a Competitive Equilibrium is achieved, where all residents have their employment demands met, as demonstrated in Figure 2(c and d) that show no variation between iterations.

The iterative nature of our method adds computational complexity to the accessibility analysis. However, the primary computational cost lies in calculating the travel cost matrix, which requires computing numerous minimum paths in the network. Since the cost matrix is calculated only once and the remaining process has negligible computational overhead, our method's overall computing time does not significantly differ from that of the traditional Balancing Time method even when 100 or more iterations are computed.

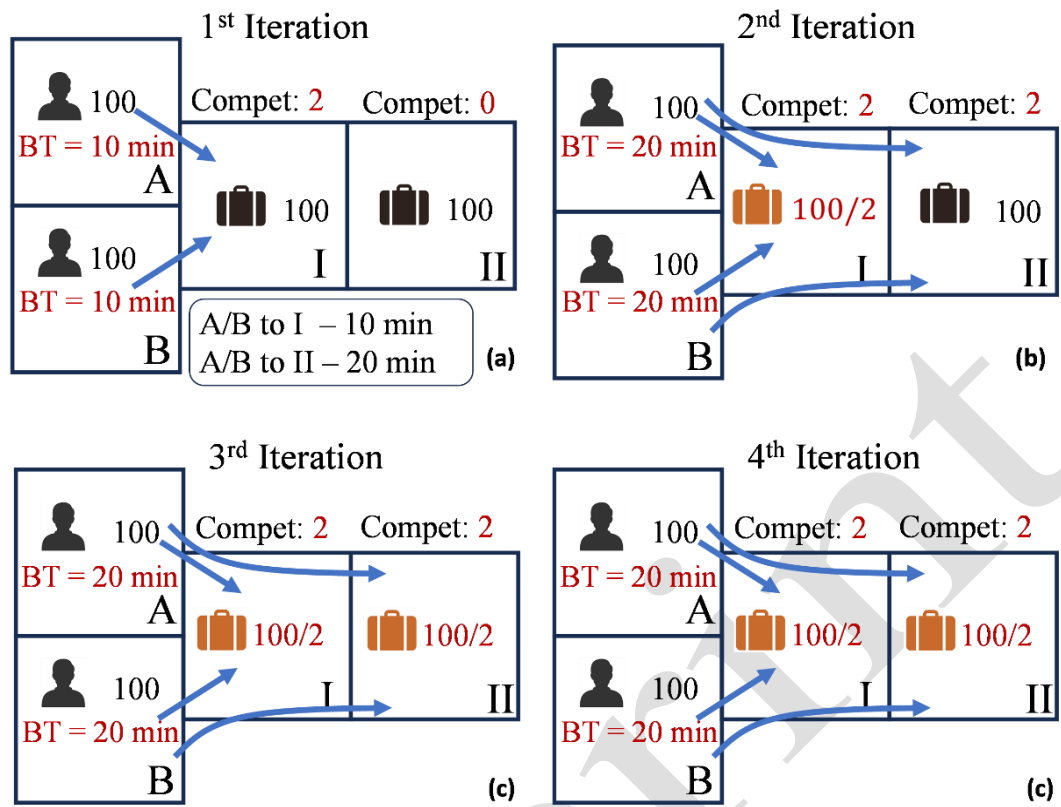

**Figure 2:** Demonstration of the Competitive Equilibrium method in three iterations. "Compet" designates the number of potential competitors in the zone in that iteration. At each iteration, the number of opportunities is divided by the number of potential competitors in the previous iteration (or maintained if compet = 0) until there is no more variation in the number of potential competitors from one iteration to the next.

Our method offers several advantages, including ease in selecting calibration parameters. Initially, the Balancing Time method has one calibration parameter: the proportion of jobs needed to saturate demand. This proportion was adopted as 1.0 by Barboza et al. (2021); that is, a job is necessary for each economically active resident. Higher values could be adopted to capture the effect of a very competitive market, trying to simulate users' search for options other than the closest ones. The most striking advantage of our method is eliminating this problem since it naturally accounts for the search for more distant opportunities in competitive markets, emerging from the iterative process.

Additional benefits include addressing problems associated with a dispersed population in small traffic zones, where Balancing Times may yield meager times. Our method also allows us to explore how job supply and demand variations can affect internal competition, which is especially important in environments with fewer jobs than economically active individuals. Furthermore, our method introduces a new metric – the number of potential competitors per zone – useful for assessing job market competitiveness and identifying regions experiencing heightened competition.

We explore these advantages in section 3 through a comparative analysis of medium-sized cities in São Paulo state, Brazil, examining how accessibility responds to different unemployment levels and market competitiveness. We focus on evaluating the distance of travel (balancing distance) in these regions, as opposed to the time to simplify the analysis for a proof of concept and avoid comparisons between modes of transport. In future applications we will consider travel time and distinctions between transportation modes in more extensive and complex networks.

**2.2. Comparative analysis of cities in São Paulo**
In this study, we compare three cities in the interior of São Paulo – São Carlos (population: 222,000), Araraquara (population: 242,000), and Rio Claro (population: 186,000) – with similar sizes and geographical positions. We obtained the transport networks for these cities as oriented graphs from OpenStreetMap using the OSMnx library in Python (Boeing, 2017). Demographic data was sourced from the 2010 census (IBGE, 2010) and redistributed into regular hexagons measuring 500 meters on each side.

To evaluate job accessibility, we relied on the geographical positions of formal jobs, which are readily available based on the postal code of enterprises from the Annual Social Information List (RAIS in Portuguese) (MTE, 2014), following a method proposed by Morelli et al. (2023). Informal job data for Brazilian cities, particularly smaller ones, are limited. Therefore, we assume that the distribution of informal jobs follows the same pattern as formal jobs. This assumption facilitates simulations of total job supply based on a simulated unemployment rate, distributing surplus jobs proportionally to the number of formal jobs.

We compare the cities on three fronts: job accessibility levels, spatial competitiveness rates, and susceptibility to variations in unemployment levels. To accomplish this, we simulate scenarios with reductions in total unemployment levels ranging from 0% to 20%, allowing us to assess how the market behaves when saturation falls below the number of economically active people in the city (opportunities/person < 1.0).

**3. RESULTS AND DISCUSSION**
This section presents the outcomes of our comparative analysis. We assess the performance of the Competitive Equilibrium method and compare it with the original Balancing Time approach proposed by Barboza et al. (2021). We achieve this by simulating various job supply scenarios, ranging from 100% saturation (one job for each economically active person) to 80% saturation (0.8 jobs per economically active person).

As an illustrative example, Figure 3 displays thematic maps depicting the balancing distances for the Balancing Time method (top row) and Competitive Equilibrium method (bottom row) for different saturation rates in São Carlos. When supply is constrained, the market intensifies competition, necessitating a metric sensitive to such variations. In this regard, Competitive Equilibrium demonstrates substantial fluctuations across scenarios, highlighting its ability to capture changes in supply. Conversely, the Balancing Time method remains relatively stable even with substantial supply restrictions. Notably, when job supply falls below demand levels, the it fails to indicate a shortage of opportunities. In contrast, Competitive Equilibrium starts revealing regions unable to meet their demands (marked in black), as would be expected when the demand exceeds supply (saturation < 1.0).

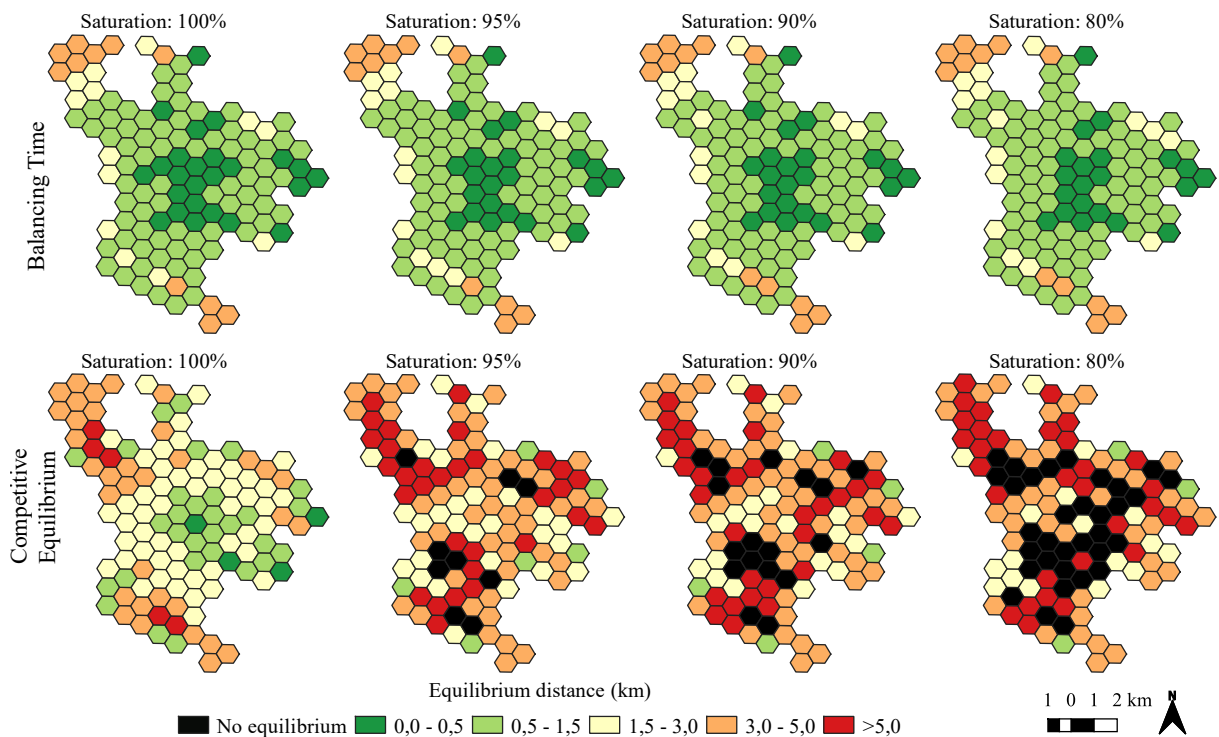

**Figure 3:** Results of the Balancing Time method (top line) and Competitive Equilibrium method (bottom line) for different levels of job supply.

Figure 4 illustrates the spatial variations in balancing distance across the three cities. When the saturation falls below 1.0, a certain proportion of the population is excluded from the job market. However, the concentration of this exclusion can vary spatially. São Carlos exhibits the highest concentration of exclusion in fewer zones. However, as shown in Figure 5, it also experiences the most significant average increase in balancing distances as competition intensifies, indicating a concentration of high competition costs in specific regions.

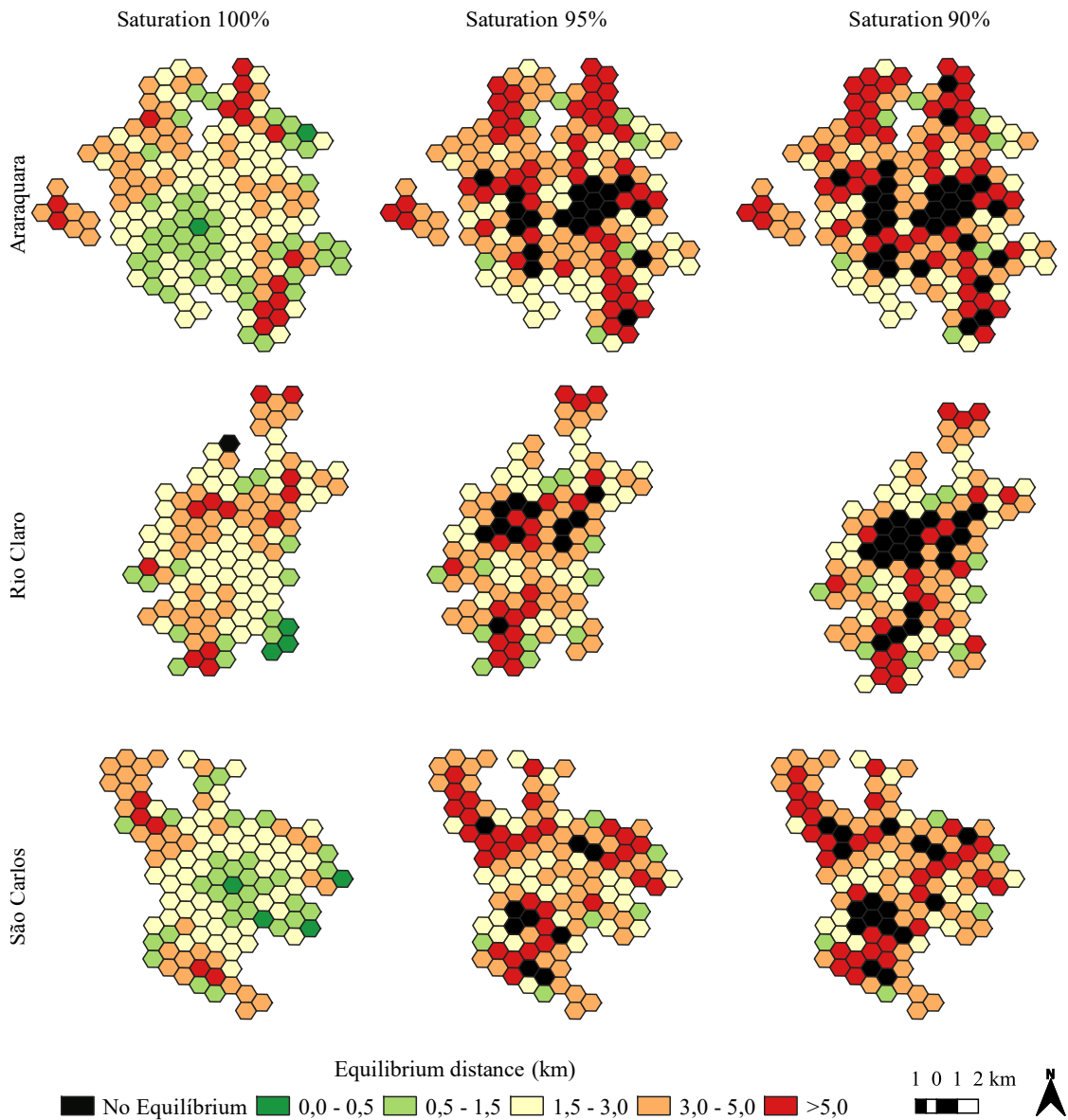

**Figure 4:** results in the three cities for three different levels of labor market saturation.

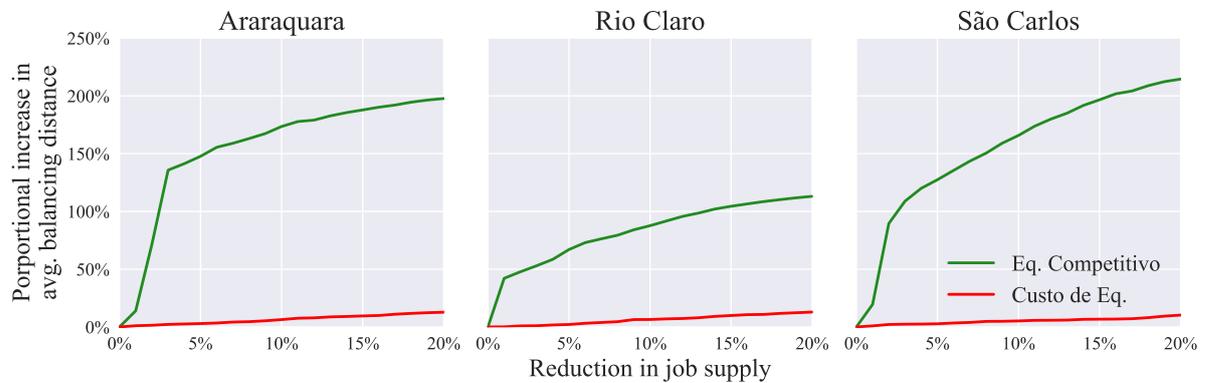

**Figure 5:** Graphs with the relative increase in the equilibrium distance.

Figure 5 provides insights into the relative changes occurring in the three cities within the database (average variation weighted by population). While balancing distances vary by less than 10% in the scenario with a 20% reduction in job supply, Competitive Equilibrium experience increases between 113% (Rio Claro) and 215% (São Carlos).

The city of Rio Claro displays smaller variations in balancing distance compared to the other cities. This difference can be attributed partly to its smaller size and compact nature, which means that even with intensified competition, individuals do not need to travel significantly longer distances to saturate the job market. São Carlos and Araraquara, on the other hand, prove less resilient to fluctuations in job supply. Consequently, these cities may encounter difficulties commuting to search for employment amid supply crises common in modern market economies.

One potential explanation for these differences lies in urban sprawl. In less compact cities, significant reductions in job supply force peripheral populations to commute longer distances to the city center, which typically has more job opportunities. As a result, sprawled cities tend to suffer more from increases in competition leading users to traverse longer distances and, as a result, increasing their reliance on motorized modes of transportation to access new opportunities. Also, the divide between the most accessible regions and the less accessible increases driven by more competition. This phenomenon is evident in Figure 6, which displays balancing distance distributions for the traditional Balancing Time method (top row) and for the Competitive Equilibrium method (bottom row). As competition intensifies, cities become increasingly divided between two categories of competitive balance: one with less impact and the other with substantial impact, as evidenced by the two distinct regions of greater thickness in the violin graphs for the Competitive Equilibrium method.

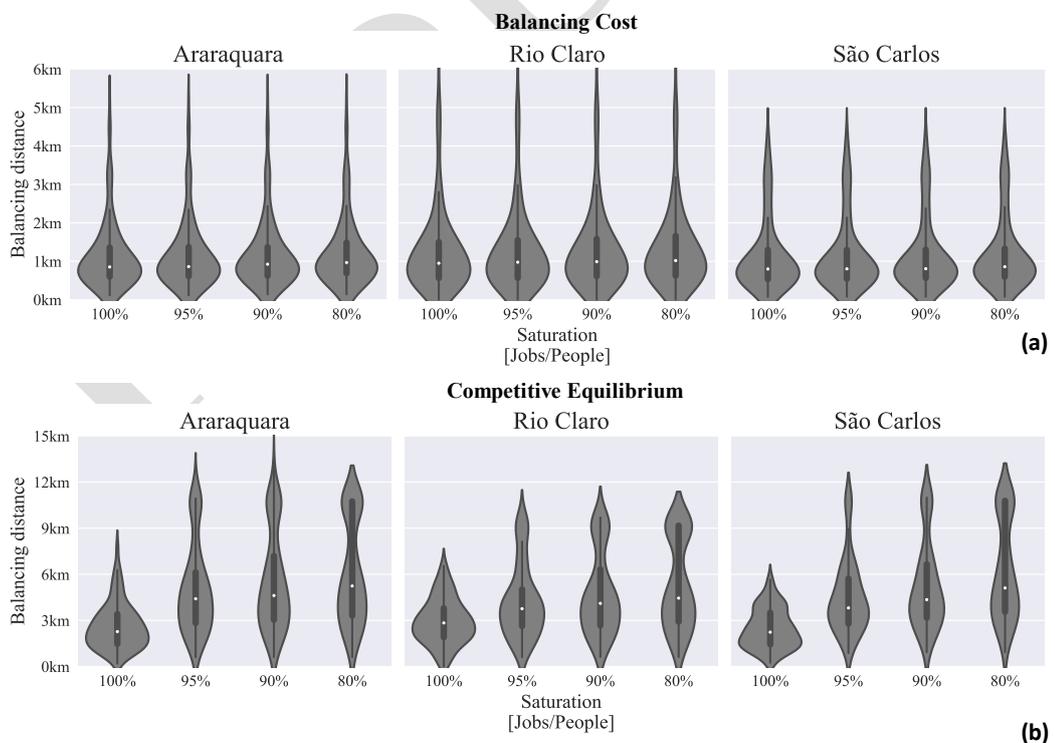

**Figure 6:** Violin plots of balancing distance distributions at different saturation levels for the Balancing Time method (a) and the Competitive Equilibrium method (b)

Figure 7 showcases the evolution in the average number of competitors per opportunity, a noteworthy result stemming from the Competitive Equilibrium method. In the first stages of reductions in saturation, São Carlos and Araraquara experience significant increases in competitors, signifying a major restructuring of their competitive landscapes, with a substantial portion of users venturing to distant locations in search of employment. Rio Claro, conversely, exhibits more modest variations. This discrepancy arises from Rio Claro's greater compactness, resulting in a naturally competitive environment, with a significant population residing near the central regions with abundant jobs. Thus, job supply reductions trigger a weaker upheaval in the city's competitive structure.

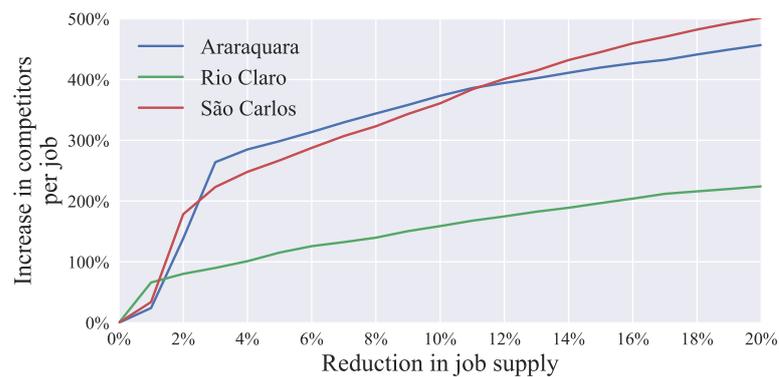

**Figure 7:** Proportional increase in the average number of competitors per job opening.

**4. CONCLUSION**

In this study, we introduced the Competitive Equilibrium method, a modification of the Balancing Time approach designed to account for competition dynamics. This method relies on an iterative process where available job opportunities are computed based on potential competitors in each region. The method has demonstrated significant sensitivity to variations in competitive markets compared to the original Balancing Time metric. Additionally, we conducted a comparative analysis in three medium-sized cities in the interior of São Paulo, Brazil.

Several advantages were identified in the use of the proposed method. Firstly, it is highly responsive to competition, making it possible to identify regions more likely to struggle to meet their market demands. This feature is particularly valuable to assess how a more abundant supply of opportunities, such as employment, can reduce the average travel distances for users. Conversely, a shortage in supply can disproportionately affect user accessibility, as observed in the cities under examination, where a 20% supply shortage resulted in transportation costs rising by 113% to 215%.

This allows for assessing how other urban opportunities may impact transportation costs for the population. For instance, increased availability of school spaces, hospital beds, green spaces in parks and leisure, and commercial facilities spread throughout cities might significantly reduce non-work-related travel and promote demand saturation. However, such evaluations need a thorough analysis of acceptable supply levels for these various types of demand, a task we plan to undertake in future research.

Applying this method also allowed us to assess the rise in the average number of competitors per employment opportunity as supply tightens. This quantitative observation enables the study of a city's propensity to reorganize its competitive landscape in response to fluctuations in supply and demand. This knowledge can be valuable for future studies seeking to identify the factors that influence increased competition with changing supply and demand conditions.

Among the cities examined, Rio Claro demonstrated lower vulnerability to job supply impacts, likely owing to its compact nature and smaller size than the other two cities. São Carlos and Araraquara exhibited similar behaviors, but a tendency to experience equilibrium distance increases approximately twice as high as Rio Claro in response to supply variations. However, more comprehensive studies involving more cities may provide clearer insights into the factors influencing a city's susceptibility to market shocks.

Finally, to ensure the replicability of our results and facilitate the application of the method in different contexts, we have made the data used in this study, along with the computational tools developed for method calculations, openly accessible. In the future, we plan to expand on the capabilities of these computational tools for more intricate analyses, particularly those involving public transportation, where additional data is essential for estimating travel times on the network.

**Funding**
This study was financed in part by the Coordenação de Aperfeiçoamento de Pessoal de Nível Superior - Brasil (CAPES) - Finance Code 001. The authors also thank the National Council for Scientific and Technological Development (CNPq) for the financial support of this study.

**Competing interests**
The authors declare that they have no known competing financial interests or personal relationships that could have appeared to influence the work reported in this paper.